\documentclass[letterpaper, 10 pt, conference]{ieeeconf}  % Comment this line out if you need a4paper

\IEEEoverridecommandlockouts                              % This command is only needed if 
\usepackage{graphicx} % for pdf, bitmapped graphics files
\usepackage{epsfig} % for postscript graphics files
\usepackage{subcaption}
\usepackage{float}
\usepackage{color}
\usepackage{wrapfig}
\usepackage{tabularx}
\usepackage{amsmath} % assumes amsmath package installed
\usepackage[hidelinks]{hyperref}

\title{\LARGE \bf
A Versatile Multi-Robot Monte Carlo Tree Search Planner \\
for On-Line Coverage Path Planning
}

\author{ Phillip Hyatt \: Zachary Brock \: Marc D. Killpack % <-this % stops a space
}

\begin{document}

\maketitle
\thispagestyle{empty}
\pagestyle{empty}

%%%%%%%%%%%%%%%%%%%%%%%%%%%%%%%%%%%%%%%%%%%%%%%%%%%%%%%%%%%%%%%%%%%%%%%%%%%%%%%%
\begin{abstract}
Mobile robots hold great promise in reducing the need for humans to perform jobs such as vacuuming, seeding, harvesting, painting, search and rescue, and inspection. In practice, these tasks must often be done without an exact map of the area and could be completed more quickly through the use of multiple robots working together. The task of simultaneously covering and mapping an area with multiple robots is known as multi-robot on-line coverage and is a growing area of research. Many multi-robot on-line coverage path planning algorithms have been developed as extensions of well established off-line coverage algorithms. In this work we introduce a novel approach to multi-robot on-line coverage path planning based on a method borrowed from game theory and machine learning - Monte Carlo Tree Search. We implement a Monte Carlo Tree Search planner and compare completion times against a Boustrophedon-based on-line multi-robot planner. The MCTS planner is shown to perform on par with the conventional Boustrophedon algorithm in simulations varying the number of robots and the density of obstacles in the map. The versatility of the MCTS planner is demonstrated by incorporating secondary objectives such as turn minimization while performing the same coverage task. The versatility of the MCTS planner suggests it is well suited to many multi-objective tasks that arise in mobile robotics.
\end{abstract}

%%%%%%%%%%%%%%%%%%%%%%%%%%%%%%%%%%%%%%%%%%%%%%%%%%%%%%%%%%%%%%%%%%%%%%%%%%%%%%%%
\section{Introduction and Motivation}
%\todo{Maybe the story of this paper is that MCTS is not just for CPP, but performs as well as a CPP algorithm. It doesn't require heuristics or pre-programmed behaviors such as lawn mowing or dividing up areas. Just maybe we show MCTS performing another task...}

Coverage Path Planning (CPP) plays a large role in the field of mobile robotics. Some examples of tasks that require a form of coverage are vacuuming, painting, seeding, harvesting, de-mining, inspection, and search and rescue. These tasks represent a large portion of work done by humans that is considered dull, dirty, or dangerous and therefore are pushing the development of CPP algorithms. While situations exist where a map is known a-priori, most of these applications require the robot to cover an area based on either partial knowledge, or no knowledge of the area. Even in relatively well mapped areas, changes in furniture layout or plant growth may cause changes in the map that require re-mapping and re-planning. In keeping with the classification established in the literature \cite{Choset2001,Galceran2013,Khan2017,Almadhoun2019}, this is referred to as on-line CPP.

The use of multiple robots for on-line CPP obviously holds significant potential when compared to single robot on-line CPP because of the benefits it presents in terms of speed and robustness to robot failures. Even for the case of a known map however, this problem has been shown to be NP hard \cite{Rekleitis2008}. Moreover, although the potential benefit increases as the number of robots grows, so does the complexity of the path planning problem. The combination of complexity with only partial knowledge of the map demands an algorithm that can reason about the most probable best path forward, taking into account the paths of other robots. This is much like the problem often faced in game theory, where the goal is to choose a probable best next action, taking into account the probable next actions of opponents.

Monte Carlo Tree Search (MCTS) is a search algorithm that is often used in decision-based game-play algorithms. It has garnered attention as being the basis for a machine learning algorithm that is a superhuman player of chess, Shogi, and Go \cite{Silver2017}. In this work, we implement MCTS as a decision based planner and use it to perform multi-robot tasks, including on-line CPP.

The goal of this paper is not to claim that our MCTS planner is necessarily the best planner for multi-robot coverage path planning tasks. Instead we intend to show that it performs similarly to a conventional planner in a coverage path planning task and that by avoiding the specification of heuristic behaviors such as lawn mowing and wall following, it is flexible enough to be used for a variety of multi-robot tasks, including those with multiple objectives.

The contributions of this paper include:

\begin{itemize}
    \item A multi-robot on-line CPP algorithm based on Monte Carlo Tree Search (MCTS)
    \item A comparison between the MCTS based algorithm and a conventional Boustrophedon-based on-line CPP algorithm
    \item A demonstration of the MCTS planner performing the same on-line CPP task with secondary objectives.
\end{itemize}

The remainder of this paper is organized as follows: Section \ref{related work} discusses previous work in the field of multi-robot on-line CPP and MCTS, Section \ref{method} explains details specific for our implementation of the Boustrophedon algorithm as well as the MCTS based planning algorithm, Section \ref{experiments} presents the simulations and experiments carried out in order to evaluate the MCTS planner, Section \ref{results and discussion} reports the results of the experiments and highlights important findings, Section \ref{conclusion} summarizes our findings and suggests future work.

\section{Related Work}\label{related work}

A great deal of the related literature for multi-robot coverage path planning has its roots in off-line and single robot coverage path planning. An efficient way to address the multiple robot path planning problem is to decompose the area into several cells, and then to cover these individual cells using established single robot algorithms \cite{karapetyan2017efficient}. Assuming all cells are reachable, and the single robot algorithms guarantee complete coverage, the entire area can be guaranteed to be covered completely.

Perhaps the simplest decomposition method uses the trapezoidal decomposition \cite{latombe2012robot,choset2005principles}, which can only handle polygonal shaped obstacles and produces many cells which can be merged together for better performance \cite{oksanen2009coverage}. The effect of merging cells from the trapezoidal decomposition is much the same as using the Boustrophedon decomposition \cite{choset1998coverage}, \cite{choset2000exact}, which itself has been shown to be a special case of the more general Morse decomposition \cite{acar2002morse}. As a comparison for our MCTS planner we choose to implement an on-line version of the Boustrophedon decomposition \cite{Rekleitis2008}.

While these cellular decomposition methods are all based on obstacle geometry, other methods are based on equal division of an area among multiple robots. These have the benefit of working in the case where the robots are already placed in the area. The work in \cite{breitenmoser2010voronoi}, \cite{cortes2004coverage}, \cite{durham2011discrete} uses the idea of Voronoi partitioning to decompose the coverage area, while in \cite{Kapoutsis2017} an optimization is used to find the decomposition which results in near-equal area for each robot to cover based on its initial position.

Once a decomposition method is chosen, either a single robot or a team of robots can be dispatched to cover the separate cells. A popular method, and the one we chose to implement for comparison with the MCTS planner, is the well known market-based method \cite{Gao2019}, \cite{dias2000market}, \cite{dias2006market}. This method allows each robot to "bid" on a task to be completed and assigns the robot with the best bid to that task. The details of the bidding are specific to each implementation.

Methods for covering cells once they are assigned are mostly derived from single-robot coverage algorithms and can vary from simple lawn-mowing algorithms \cite{lumelsky1990dynamic}, \cite{choset1998coverage}, \cite{acar2002morse} to the more complex wavefront algorithm \cite{zelinsky1993planning} to the popular Spanning Tree Coverage (STC) method \cite{gabriely2001spanning}. For our Boustrophedon planner used to compare with the MCTS planner, we implement a lawn mowing algorithm which includes preliminary wall following.

Notably, the STC algorithm \cite{gabriely2001spanning} and several of its multi-robot variants \cite{agmon2006constructing}, \cite{hazon2005redundancy}, \cite{zheng2005multi} do not depend on a cellular decomposition, but are classified as off-line methods because they depend on knowledge of the map a-priori in order to build a spanning tree. In \cite{choi2009online}, \cite{lee2011smooth} on-line variations are derived by allowing the robots to build a spanning tree as they spiral outward from their starting locations. Other on-line methods which do not depend on cellular decomposition are those that are based on neural networks \cite{luo2002solution}, \cite{yang2004neural}, \cite{luo2008bioinspired} and insect behaviors \cite{Wagner2008}. Our MCTS planner is similar to these methods in that it does not use cellular decomposition and can still be on-line for unknown environments, however it differs in that it uses a tree search to find robot paths. 
%The flexibility afforded by using a tree search method means that the same planner can be used to perform multiple tasks by changing only the objective or objective function.

Monte Carlo Tree Search was first used in the field of game theory as an AI for board games and video games \cite{Chaslot2008}. Since then it has been used in several domains, including the field of multi-agent planning for active perception tasks \cite{Best2019}, \cite{Best2016} and exploration \cite{Corah2017}, \cite{Arora2017}. Because MCTS is essentially performing an optimization over possible actions, it is very versatile. As will be shown in Section \ref{experiments}, only the objective function used in the tree search must be modified in order to complete several different tasks. For the broad field of coverage path planning this versatility is attractive because of the different priorities to be optimized such as minimizing energy \cite{DiFranco2015}, \cite{Wei2018} or minimizing the number of turns \cite{Vandermeulen2019}, \cite{Bochkarev2016}. Although not implemented in this work, MCTS can also be used to model uncertainty as in \cite{VanPham2018}, \cite{Li2019}.

An active area of MCTS research related to our implementation is MCTS parallelization \cite{Rocki2011} \cite{Browne2012}. In \cite{Chaslot2008parallelization} the authors present three methods of parallelization which are meant to increase the speed of convergence of the MCTS algorithm. While our method of parallelization is most similar to the method referred to as "root parallelization", where several entire Monte Carlo Trees are explored in parallel, a key difference is that the parallel Monte Carlo Trees in our implementation do not represent the same robot, but each tree represents a separate robot.

While communication between robots is a concern in the field of mobile robotics (\cite{Best2018}, \cite{Rekleitis2004}), for simplicity in this work we assume that all robots have uninterrupted communication between each other.

\section{Methods}\label{method}
%In this section we describe the two algorithms that were used to accomplish the multi-robot on-line coverage tasks in the Experiments section.

% For example:

% In this section, we first outline our implementation of the Boustrophedon multi-robot coverage algorithm described in \cite{Rekleitis2008}. The MCTS-based algorithm is then discussed. We chose to implement the Boustrophedon algorithm as a comparison point for the MCTS method because it appeared from our review to have the fewest limitations based on map discretization, obstacle shape and size, and sensor range when compared with other on-line methods.

\subsection{Multi-Robot Boustrophedon Coverage Path Planner}

Our Boustrophedon planner is based on the work found in \cite{Rekleitis2008} on multi-robot Boustrophedon coverage. We first give a brief overview of the algorithm, then point out details specific to our implementation.

As described in Section 5 of \cite{Rekleitis2008}, the Boustrophedon algorithm begins by dividing the area into cells. Since no obstacles are known at first, the cells begin as stripes of the area to be covered. Robots each begin covering their cells using a wall following algorithm, following either the virtual cell walls or real obstacles until the start point within their cell is reached. As obstacles are detected and mapped, new cells are formed and added to the adjacency (or Reeb) graph. Robots perform complete coverage of their assigned cells using a lawn mowing algorithm.
%This is analogous to the path an ox would take while plowing a field - hence the name "Boustrophedon", which is derived from a Greek term meaning "of the ox" - or that someone might take while mowing a lawn. By dividing the area into cells this way and always returning to the original point, the robots can infer that their coverage of each cell is complete. 
%As the robots cover each cell, they log the position of any obstacles encountered at the cell boundaries; once coverage of each cell is completed, these obstacle locations are used to estimate the location and dimensions of cells to either side of the completed cell. These estimated cells are added to a queue of unassigned cells, which the robots query upon finishing each cell to identify their next assignment.

Again, we assume that robots share map and coverage information without restriction throughout the simulation. When a robot completes coverage of an assigned cell, the robot is given a new assignment from a list of unassigned cells taken from the adjacency graph. A centralized computer or master robot is utilized to handle management of the adjacency graph, task allocation, and map synchronization between robots. The method of task allocation uses the market-based approach, where each robot's bid for an unassigned cell is a function of distance to the cell and completion of their current cell. To plan robot paths between cells, we use the wavefront algorithm \cite{zelinsky1993planning}.

A key feature of our implementation which differentiates it from the original work is that we divide the entire area into a grid of squares that are the same width as the robot. We will refer to the cells formed using the Boustrophedon decomposition as "cells", and the robot-sized squares that make up the grid of the entire area as "grid squares".

%Our implementation of this algorithm sought to follow the description outlined in \cite{Rekleitis2008} as closely as possible; however, to provide a more equivalent test case to the MCTS algorithm described in this paper, the algorithm was modified to discretize the space into a grid of robot-sized squares. 

% To simulate the algorithm, each robot keeps the following information stored locally:

% \begin{enumerate}
% %  \item The dimensions of the map
%   \item Local obstacle and coverage maps that are updated as the robots sense and cover the area
%   \item Starting location of coverage, updated for each cell the robot covers
%   \item The robot's current mode (Idle, Travelling to Cell, or Covering Cell)
%   \item Temporary lists of encountered obstacles' coordinates to help with new cell estimation
%  \end{enumerate}

%The robots function in a largely autonomous manner during their atomic tasks of coverage and travel between cells. The main mode of communication between robots occurs at the completion of a cell; whenever a robot completes its cell, it informs the other robots that the cell is complete and provides them with any new cell estimates that it was able to encounter. At this time, the robots' coverage and obstacle maps are also synchronized. This allows the robots to have the best information about the map available prior to beginning travel to their new assignments.

\subsection{Multi-Robot Monte Carlo Tree Search Planner}
Because MCTS has been shown to effectively find good decisions in a very large decision space, we implemented it as the basis for a sampling-based planner for multiple robots. We aim to show that this planner performs comparably to a conventional planner for a coverage path planning task, while maintaining the flexibility to incorporate secondary objectives or even be used for other tasks. In this section we give a brief description of MCTS for the purpose of explaining our algorithm, however a more in-depth resource for understanding MCTS can be found in \cite{Browne2012}.

\begin{figure}
    \centering
    \includegraphics[width=0.5\textwidth]{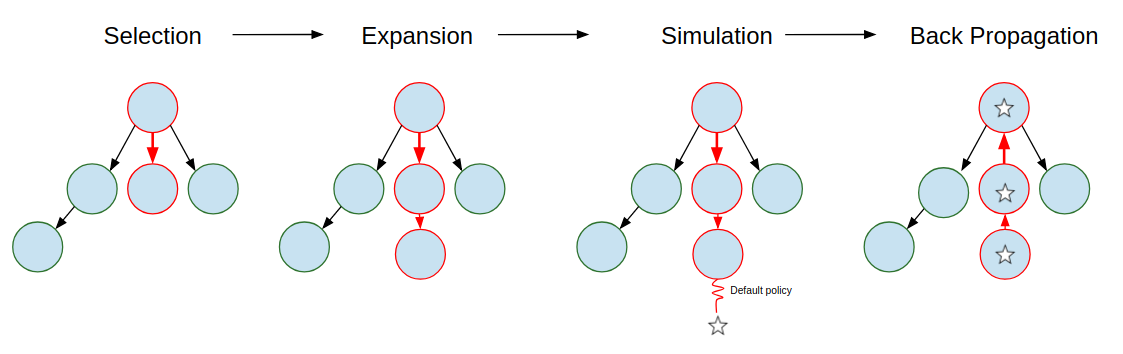}
    \caption{Visual representation of Monte Carlo Tree Search Algorithm}
    \label{fig:mcts_diagram}
\end{figure}

A visual representation of the MCTS algorithm can be seen in Figure \ref{fig:mcts_diagram}. For this work, nodes in the tree represent robot states and arrows represent actions. Each robot has its own tree, with the root node representing the current state of the robot. We assume that in any state a robot can only choose three actions (to move left, right, or straight one grid square). 

The idea behind the MCTS planner is to grow the search tree as far as possible, forming the best possible estimate of the value of each state adjacent to the root node, until an action must be taken. At that point, the action which leads to the highest value state is taken, the robot's new state becomes the root node, and the process is repeated, possibly with new information about the environment.

In order to grow the search tree until an action must be taken, the following four steps are repeated, starting from the root of the tree:
\begin{enumerate}
    \item \textit{Selection: } Calculate the Upper Confidence Bound for Trees (UCT) associated with each child of the current node. UCT is calculated using the following equation:  
    
    \begin{equation}
    \label{uct_eqn}
        UCT = X + 2C_{p} \sqrt{\frac{2\ln{n}}{n_{j}}}
    \end{equation}
    
    where $X$ is the node's value (assigned in step 4), $n_j$ is the number of times it has been visited since creation of the tree, $n$ is the number of times its parent has been visited, and $C_p > 0$ is an exploration coefficient. 
    %As can be seen from Eqn. \ref{uct_eqn}, high values of $C_p$ encourage exploration, while low values encourage exploitation. 
    Note that in an unvisited node $n_j=0$ therefore $UCT=\infty$, meaning all children of a parent node will be visited before any child is visited twice. Iterate calculating UCT for nodes and moving to the node with the highest UCT until you reach a leaf node (a node with no visited children).  
    
    \item \textit{Expansion: } Expand this node by randomly choosing one of the un-simulated actions
    
    \item \textit{Simulation: } Forward simulate the system based on a default policy until the end of a defined time horizon. Calculate a value ($X$) associated with this action based on evaluation of an objective function along the trajectory from the root node.
    
    \item \textit{Backpropagation: } Assign the value $X$ to the newly created node. Backpropagate this value all the way back to the root of the tree. The value of each node is simply the average of the values of its children.
\end{enumerate}

Instead of considering the entire multi-robot system as one search tree, we choose to treat each robot as having its own individual search tree. This allows us to parallelize the MCTS algorithm and greatly increase the speed and depth of the search in a given amount of time. This also could allow this algorithm to be run as either a centralized or decentralized planner. The only information that robots must share with each other is their discovered map, their best path at the last solve, and in the case of CPP, their covered grid squares. 

%The initial state (root) of each tree is simply the starting pose of the robot, and the action space consists of three actions - move straight one grid square, move left one grid square, or move right one grid square. 

During the simulation phase of our MCTS planner, for each time step, each robot first simulates the actions of all other robots based on their last known best path. This allows the planner to take into account the movement of other robots. The robot then simulates their own action and accumulates the value calculated by the objective function for this state and action.

%Not surprisingly, the performance of the MCTS algorithm depends heavily on the selected objective function. In fact, while the focus of this work is on a team of robots mapping and covering an area, by using a different default policy and objective function, this method could be used for any multi-robot task.

The default policy used for this work is to move in a straight line until the grid square straight ahead is either occupied, or has already been covered. Once one of these conditions is reached, the robot will look right and left for grid squares that are not occupied or have not been covered. If only one of the grid squares is not occupied or has not been covered, the robot will turn toward that grid square. In the case where neither of the grid squares are occupied or have been covered and the grid square straight ahead is occupied, the robot randomly chooses to turn left or right. If both grid squares have been covered and the grid square ahead is unoccupied, the robot continues straight.
%For performing non-coverage tasks, every grid square in the map is considered covered.

A semi-intelligent default policy was deemed necessary in our planner because of the possibility for the robot to get stuck hitting obstacles. Without a default policy which avoided obstacles, the planner found very little incentive to explore areas which contained many obstacles. Because the default policy does seem to have an effect on solutions, future work could include experimenting with different default policies or even using a learned method such as a Deep Neural Network as a default policy.

However, the objective function has the greatest affect on the multi-robot team behavior. As will be shown in Section \ref{experiments}, leaving all other parts of the algorithm constant and changing the objective function completely changes the behavior of the MCTS planner. All objective functions for this work are evaluated over a robot's trajectory throughout a simulation of length $T$ time steps.

For the experiments in section \ref{mcts_vs_Boustrophedon} comparing the MCTS planner to the Boustrophedon planner, the value ($X$) was defined as

\begin{equation}
    X = \sum_{k=1}^T \bigg[\frac{1}{(t_k+1)^2}(p_k - C_{hit}q_k)\bigg]
    \label{eq:coverage_value_function}
\end{equation}

where

\begin{equation}
  \begin{cases} 
      p_k = 1 & $if robot "covered" a grid square at time step $ k \\
      p_k = 0 & $otherwise$
   \end{cases}
\end{equation}

\begin{equation}
   \begin{cases}
      q_k = 1 & $if robot hit a wall or robot at time step $ k \\
      q_k = 0 & $otherwise$
   \end{cases}
 \end{equation}

$t_k$ is time in seconds since the beginning of the simulation, and $T$ is the final time step of the simulation horizon. Note that time step $k=0$ represents the initial time before any actions have been taken, therefore penalizing or rewarding at $k=0$ would only add a constant offset to every value.

The value function used for the experiments in section \ref{multiple_objectives} with a secondary objective was

\begin{equation}
    X = \sum_{k=1}^T \bigg[\frac{1}{(t_k+1)^2}(p_k - C_{hit}q_k - C_{turn}r_k)\bigg]  
    \label{eq:multi_objective_value_function}
\end{equation}

where

 \begin{equation}
   \begin{cases}
      r_k = 1 & $if robot turned left/right/either at time step $ k\\
      r_k = 0 & $otherwise$.
    \end{cases}
\end{equation}
In essence, the first objective function rewards the robot for covering grid squares. The decay of the reward over time reflects the fact that values calculated farther in the future are less certain. 

The second objective function is the same, with the addition of a cost on turning either left, right, or either. As will be shown in section \ref{multiple_objectives}, this simple addition to the cost function has a significant effect on robot behavior.

The values of $C_p$ and $T$ are tuning parameters that can be changed to alter the performance of MCTS. Higher values of $C_p$ encourage more exploration, while lower values encourage exploitation. While increasing $T$ allows MCTS to look farther into the future, it also takes longer to perform the simulation phase of MCTS and therefore does not allow for deep exploration into the search tree. For this work we used $C_p=1.0$, $T=30$, and $C_{hit}=2.0$. 
%In simple experiments omitted from this work for brevity, it was found that the quantitative performance of the MCTS planning algorithm was not very sensitive to these tuning parameters. Future work may include a more in-depth exploration of the effects of these tuning parameters as well as the effects of changing default policy and objective function.

\section{Experiments}\label{experiments}
In this section we describe the experiments carried out in order to compare the MCTS planner to the Boustrophedon planner. We study their performance in problems of varying complexity by varying the number of robots as well as the number of obstacles. We also describe the experiments which highlight the versatility of the MCTS planner by introducing secondary objectives.

All experiments were carried out in a real-time simulation on an Intel i7-4770 processor using C++ and the Robot Operating System (ROS). We assumed that robots could localize perfectly and that each was equipped with a perfect omnidirectional range sensor for mapping. Robots were assumed to have a width of 0.5 m and therefore the map was divided into 0.5 m grid squares. Robots were assumed to travel at a constant velocity of 1 m/s and to turn instantaneously. While no actual ground robot can turn instantaneously, a smoothing function could be used to turn sharp corners in paths into arcs. These paths would cover a similar area with little need for robot deceleration. Alternatively, minimizing turns could be made a secondary objective as in Section \ref{multiple_objectives}.

The fact that the simulation was real-time especially effected the MCTS planner because it continuously iterates through the four MCTS steps outlined above until the solve time is reached. We required the planners to re-plan every 0.5 seconds, which corresponds to the time it takes to for a robot travel from the center of one grid square to the center of another grid square.

In order to make the simulation more similar to a real-world task with an unknown map, the robots initially knew nothing about the map except its size. As each robot discovered parts of the map, they shared this information with all other robots. The sensors provided information only about grid squares which were within line of sight of the robots and within the sensor range. The sensor range was limited to two meters, similar to that of an inexpensive infrared or ultrasonic range sensor.
%The effect of varying sensor ranges is explored in one of the experiments.

Maps were all randomly-generated 10 m square maps with obstacles placed anywhere except the map border. In the case where randomly distributed obstacles encircled grid squares (making them inaccessible to the robots), the inner grid squares were manually filled in. The effect of varying the density of obstacles is explored in section \ref{obstacle_density_experiments}.

Because the simulation and algorithms are not completely deterministic, statistics over many trials are represented using box and whisker plots. For this work, the box edges represent the first and third quartile of the results, the center line represents the median, and the whiskers represent samples inside 1.5 times the interquartile range (the distance between the first and third quartile).
%values not considered outliers (an outlier is a value that is more than 1.5 times the interquartile range away from the 25th or 75th percentile)\todoMarc{what do you mean not considered outliers, aren't extreme values by definition outliers? Can you clarify this a bit?} 

\subsection{MCTS Planner vs Boustrophedon Planner}
\label{mcts_vs_Boustrophedon}
%\todo{The optimality ratio just shows that neither of our path planners are garbage...if people know what a good optimality ratio is. Should we just report completion time instead?}
%In order to quantify performance for the coverage task, we report the ratio of completion time to optimal completion time instead of reporting completion time itself. This performance metric was established in \cite{zheng2005multi} and is agnostic to map size, obstacle density, and number of robots. It should be noted that while a ratio of 1.0 would be optimal, this score is likely impossible for many maps and initial robot positions. For the sake of clarity, we define our optimality ratio ($Ratio$):

% \begin{equation}
% Ratio = t_{completion}/t_{optimal}.
% \end{equation}

% where optimal completion time is calculated as:

% \begin{equation}
% t_{optimal} = \frac{number\ of\ grid squares\ to\ be\ covered}{(robot\ speed)(number\ of\ robots)}
% \end{equation}

% and $robot\ speed$ is in units of grid squares per second per robot.

 \begin{figure}[h]
    \centering
     \subcaptionbox{}{\includegraphics[width=1.5cm]{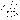}}
     \subcaptionbox{}{\includegraphics[width=1.5cm]{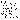}}
     \subcaptionbox{}{\includegraphics[width=1.5cm]{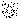}}
     \subcaptionbox{}{\includegraphics[width=1.5cm]{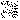}}
 \caption{Examples of maps with varying obstacle densities. a) 5 \% obstacle density, b) 10 \% percent obstacle density, c) 15 \% obstacle density, d) 20 \% obstacle density} 
 \label{fig:final_optimality_vs_number_of_robots}
 \end{figure}

The Boustrophedon planner requires that all robots be uniformly spaced at one edge of the map for initialization, however the MCTS planner does not. For a direct comparison, we run experiments on the MCTS planner for the case in which the robots start in the same places as the Boustrophedon planner. To highlight the versatility of the MCTS planner, we also run experiments on the MCTS planner in which the robots are initialized in random locations and orientations. These positions and orientations are both drawn from uniform random distributions.
%To ensure that each robot had a relatively equal number of two grid square width stripes to cover in the Boustrophedon algorithm, the map's width in grid squares was divided by two to obtain the number of stripes in the map. This stripe count was then divided by the number of robots to determine the inter-robot spacing, leaving any remaining grid square columns to the very right of the map. 
%This method was also used for the MCTS algorithm to ensure that our comparisons were not affected by changing the initial robot locations, however in experiments omitted for brevity it was found that MCTS planner performance was similar regardless of starting location.

\subsubsection{Varying Number of Robots}

\begin{figure}[h]
    \centering
     \includegraphics[width=7cm]{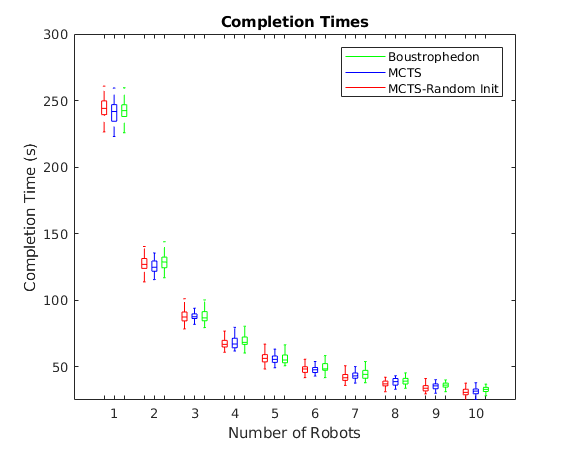}
 \caption{Completion time vs. Number of Robots for the MCTS and Boustrophedon multi-robot on-line CPP algorithms. MCTS-Random Init shows results for when the robots are initialized at random locations in the map.} 
 \label{fig:completion_time_vs_number_of_robots}
 \end{figure}

The first experiment was designed to evaluate how the multi-robot on-line CPP algorithms performed with increasing number of robots. Using five 10 \% obstacle density maps, simulations were run for teams of one to ten robots. Ten simulations were run on each map for each planner. For each planner, the statistics reported are over all fifty trials over all five maps. The total time to cover every grid square is reported in Figure \ref{fig:completion_time_vs_number_of_robots}.

%Both coverage time and the ratio of coverage time to the optimal coverage time are reported in Figures \ref{fig:completion_time_vs_number_of_robots} and \ref{fig:final_optimality_vs_number_of_robots}.

%   \begin{figure}[h]
%     \centering
%     %\includegraphics[width=7cm]{figures/final_optimality_vs_number_of_robots.jpg}
%     \includegraphics[width=7cm]{figures/new_mcts_optimality_ratio.png}
%  \caption{Ratio of Completion time to Optimal Completion time vs. Number of Robots for both the MCTS and Boustrophedon multi-robot on-line CPP algorithms. MCTS-Random Init shows results for when the robots are initialized at random locations in the map.} 
%  \label{fig:final_optimality_vs_number_of_robots}
%  \vspace{-0.5cm}
%  \end{figure}

\subsubsection{Varying Obstacle Density} \label{obstacle_density_experiments}
\begin{figure}[h]
    \centering
     \includegraphics[width=7cm]{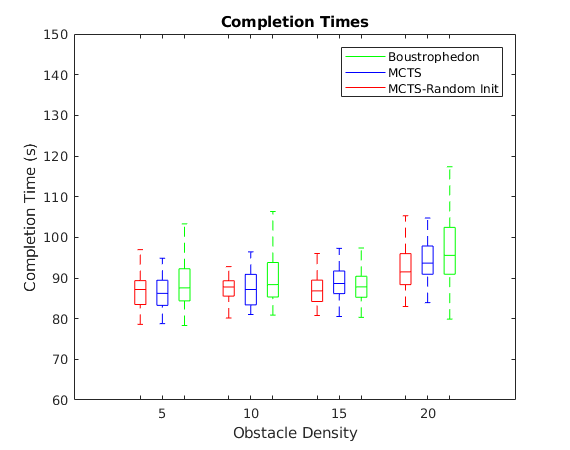}
 \caption{Comparison of completion times for a coverage task of the same size area with varying obstacle density.} 
 \label{fig:completion_time_vs_obstacle_density}
 \end{figure}

The second experiment was designed to evaluate how the multi-robot on-line CPP algorithms performed with increasing obstacle density in the map. Using teams of 3 robots and a sensor range of 2 m, simulations were run on maps with obstacle densities of 5, 10, 15, and 20 percent. Five maps of each density were generated. Ten simulations were run on each map for each planner. For each planner, the statistics reported are over all fifty trials over all five maps. The coverage times are reported in Figure \ref{fig:completion_time_vs_obstacle_density}.

%  \begin{figure}[h]
%     \centering
%     %\includegraphics[width=7cm]{figures/new_mcts_obstacle_density_optimality_ratio.png}
%     \includegraphics[width=7cm]{figures/new_mcts_obstacle_density_completion_times.png}
%  \caption{Ratio of Completion time to Optimal Completion time vs. Obstacle Density for both the MCTS and Boustrophedon multi-robot on-line CPP algorithms} 
%  \label{fig:final_optimality_vs_obstacle_density}
%  \vspace{-0.5cm}
%  \end{figure}

% \subsubsection{Varying Sensor Range}
% The last experiment was designed to evaluate how the multi-robot on-line CPP algorithms performed with increasing sensor range. Using teams of 3 robots and maps with obstacle density of 10 \%, simulations were run with sensor ranges of 2m, 3m, 4m, and 5m. Ten simulations were run on each map for each planner. For each planner, the statistics reported are over all fifty trials over all five maps. The ratio of coverage time to the optimal coverage time are reported and can be seen in Figure \ref{fig:final_optimality_vs_sensor_range}.

% \begin{figure}[h]
%     \centering
%      \includegraphics[width=7cm]{figures/completion_time_vs_sensor_range.jpg}
%  \caption{caption} 
%  \label{fig:completion_time_vs_sensor_range}
%  \end{figure}
 
%  \begin{figure}[h]
%     \centering
%      \includegraphics[width=7cm]{figures/final_optimality_vs_sensor_range.jpg}
%  \caption{Ratio of Completion time to Optimal Completion time vs. Sensor Range for both the MCTS and Boustrophedon multi-robot on-line CPP algorithms} 
%  \label{fig:final_optimality_vs_sensor_range}
%  \vspace{-0.5cm}
%  \end{figure}

\subsection{MCTS Planner with Secondary Objectives}
\label{multiple_objectives}
In order to demonstrate the versatility of the MCTS planner, we introduce secondary objectives to the CPP problem. Note that by using a more complex model for simulation, objectives such as minimizing energy or time spent out of communication range could be implemented. Using our simple robot model however, we chose the secondary objectives of minimizing left turns, minimizing right turns, and minimizing any turns.

In order to incorporate these secondary objectives, we use the objective function defined in Equation \ref{eq:multi_objective_value_function}. For the case of minimizing left or minimizing right turns $C_{turn}=.5$, while for the objective of minimizing any turns we use $C_{turn}=.1$. For comparison, we compare these solutions to those with no cost on turns ($C_{turn}=0$).

\begin{figure}[h]
    \centering
     \includegraphics[width=7cm]{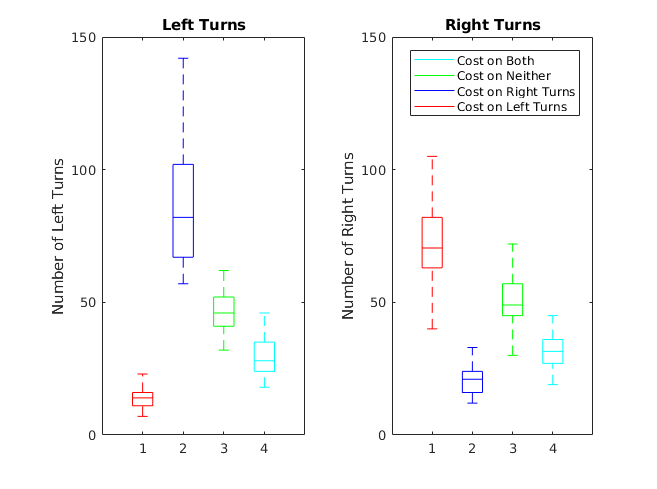}
 \caption{Number of left and right turns using the MCTS planner with four different costs. Task completion times for these experiments are reported in Figure \ref{fig:turn_minimizing_completion_times}.} 
 \label{fig:left_and_right_turns}
 \end{figure}

\begin{figure}[h]
    \centering
     \includegraphics[width=7cm]{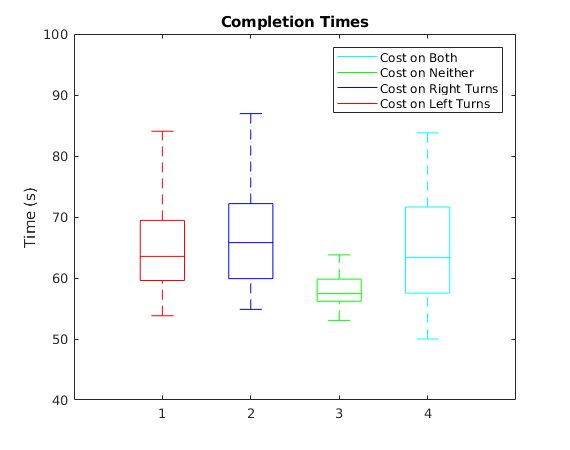}
 \caption{Completion times using the MCTS planner with four different costs.}
 \label{fig:turn_minimizing_completion_times}
 \end{figure}

Experiments were carried out with teams of five robots on maps with 10\% obstacle density. Figure \ref{fig:left_and_right_turns} shows the resulting number of turns for each slightly modified objective function. The number of left turns is shown on the left, the number of right turns on the right, and each objective function is represented by a different color. The completion time for each is represented in Figure \ref{fig:turn_minimizing_completion_times}.

\section{Results and Discussion}\label{results and discussion}
This section contains quantitative results. A qualitative example of the MCTS algorithm is also available at \url{https://youtu.be/ueyKdxn_07Y}.

\subsection{MCTS Planner vs Boustrophedon Planner}
\subsubsection{Varying Number of Robots}
The completion times from the trials with varied number of robots are shown in Figure \ref{fig:completion_time_vs_number_of_robots}. The completion time holds an inverse exponential relationship with the number of teamed robots for both the Boustrophedon and MCTS algorithms, which is the expected result; more robots will cover an area more quickly, with smaller time gains accrued for each additional robot \cite{zheng2005multi}. 
%This result also speaks to the increased ratio to optimal completion time (as described in ) also seen with increasing number of robots, as shown in Figure \ref{fig:final_optimality_vs_number_of_robots}. 
As the number of robots increases, there is a corresponding increase in potential for repeat coverage or robot idling due to imperfect knowledge of the coverage area.

For more than three robots, the MCTS planner completes the tasks slightly faster than the Boustrophedon planner, while for three or fewer robots the two methods are very comparable. This suggests that despite looking for good robot actions over a very large search space, the MCTS planner is capable of finding good solutions quickly.

On average, the MCTS planner with randomly placed robots completes the tasks in the least time. This is likely because the randomly placed robots are spaced more evenly over the workspace than those spaced evenly along one wall. This allows for faster discovery of obstacles and also decreases the likelihood of repeat coverage since the robots are farther from each other. These results indicate that where it is possible to deploy the mobile robots throughout the area more evenly, coverage tasks can be completed more quickly. While the MCTS planner's versatility allows for this, the Boustrophedon algorithm does not.

\subsubsection{Varying Obstacle Density}
Completion times for the obstacle density experiments are shown in Figure \ref{fig:completion_time_vs_obstacle_density}. As can be seen from the figure, increasing obstacle density only very slightly, if at all, increases the time to completion. An increase in completion time is expected because the robots must navigate around obstacles and repeat coverage in small corridors with access to more open areas. However, this is partly offset by the fact that the increase in obstacles means that there are less cells to cover. The performance of the two planners is comparable and close to the performance observed in Figure \ref{fig:completion_time_vs_number_of_robots}, demonstrating that the completion time for both planners is almost unaffected by obstacle density.

\subsection{MCTS Planner with Secondary Objectives}
Figure \ref{fig:left_and_right_turns} shows the results of the experiments with secondary objectives. The results for the planner with no cost on turning (green) serve as a baseline for comparison. We can see that by adding a cost on left turns, the number of left turns decreased, while the number of right turns increased. When the cost is on right turns, we observe the opposite. By placing a cost on both left and right turns, both decrease. The trends represented in Figure \ref{fig:left_and_right_turns} demonstrate that the MCTS planner is capable of realizing secondary objectives while coverage path planning without the need for implementation of a new planner or modifying heuristic behaviors such as lawn mowing or wall following. The only modification necessary was in the objective function.

In Figure \ref{fig:turn_minimizing_completion_times} we see the completion times for the four different tasks with secondary objectives. We see that the lowest completion times are accomplished by the MCTS planner with no cost on turning. This is because the objectives of minimizing turns and covering grid squares are competing objectives. As with any multi-objective optimization, there exists a continuum of solutions (or pareto front) which represent the trade-off between two or more competing objectives. The solutions shown here are just a couple of points along that Pareto front, however by adjusting the relative weightings it is possible to find solutions with fewer turns at the cost of longer completion times. The ability to tune the MCTS planner for desired behavior using only a simple objective function is one of the main strengths of the MCTS approach.

\section{Conclusion and Future Work}\label{conclusion}
In this paper we have presented a novel MCTS-based multi-robot on-line coverage path planner. We have shown that this MCTS planner performs just as well as the Boustrophedon planner for small numbers of robots, while performing slightly better for larger numbers of robots. The added versatility of the MCTS planner allows for more uniform spacing of the robots over the coverage area leading to further improvement, meaning a decrease in coverage time.

Aside from the slightly improved performance over the Boustrophedon planner, the strength of the MCTS algorithm lies in its versatility. The MCTS algorithm is approximating the value of possible actions based on an objective function defined by the user and a simulation. The simulation can be of any fidelity and can even incorporate stochastic effects if desired. These are flexibilities not afforded by the Boustrophedon planner or other heuristically based planners. 
%Also, while the Boustrophedon algorithm required the robots to start equally spaced on one side of the map, the MCTS-based algorithm is able to initialize robots anywhere in the map with no performance decrease.

In addition, while typical CPP algorithms have very little flexibility for changing robot behavior, the MCTS-based algorithm has several ways to change behaviors. As demonstrated, the objective function used in the MCTS planner can be changed to punish or reward certain behaviors. In fact, by modifying only the objective function this same planner was shown to perform a target following task for multiple robots in cluttered environments. While results were not quantified for this paper, qualitative results can be seen in the video: \url{https://youtu.be/ueyKdxn_07Y}. Future work may explore the effects which default policy, horizon length, and exploration coefficient may have on behavior.

As multi-robot CPP tasks become more widely used in several domains, especially in unknown environments, we believe that versatile CPP algorithms such as the presented MCTS planner will be of increasing value. 

%It should be noted that the default policy used in the MCTS planner was also very simple and could be easily replaced by a more sophisticated algorithm. This would help to improve the planner's approximation of the value of each choice. Future work may include using a non-sampling based algorithm such as the Boustrophedon planner, a neural network based planner such as in \cite{ChaominLuo2002} or even a Deep Neural Network as a default policy.

% \include{references}
\bibliography{references}
\bibliographystyle{IEEEtran}

\end{document}